# Organizational transformation: The impact of servant leadership on work ethic culture with burnout as a mediating factor in the hospitality industry

**Authors:**
Darul Wiyono[1]
Rinaldi Tanjung[2]
Hedi Setiadi[3]
Sri Marini[4]
Yayan Sugiarto[5]

**Affiliations:**
[1,2,3]Management Administration Study Program, Ariyanti Academy of Secretary and Management, Bandung, Indonesia

[4,5]Tourism Academy NHI's Hospitality Program in Bandung, West Java, Indonesia

**Corresponding author:**
Darul Wiyono,
darulwiyono96@gmail.com

**Orientation:** The study explores the connections among servant leadership, burnout, and work ethic culture in organizations. It aims to provide a detailed understanding of how servant leadership influences work ethic culture, especially by considering the role of burnout.

**Research Purpose:** This study aims to understand how servant leadership influences work ethic culture and explore the mediating role of burnout in this relationship.

**Motivation for the Study:** This study wants to fill gaps in our understanding of how servant leadership, burnout, and work ethic culture are connected. It seeks to add useful insights to what we already know from previous research.

**Research Approach/Design and Method:** The study, using surveys and statistics, examines the links between servant leadership, burnout, and work ethic culture in 113 hotels in Bandung, Indonesia, with 339 participants. A 183-sample, chosen with a 0.05 margin of error, underwent SEM-PLS analysis using SmartPLS 3.0.

**Main Findings:** The key findings underscore that servant leadership exerts a positive influence on work ethic culture, and burnout plays a pivotal mediating role in this dynamic. The results shed light on the intricate dynamics shaping organizational cultures.

**Practical/Managerial Implications:** The findings aid organizations in forming supportive leadership policies, promoting employee well-being, and fostering ethical work culture. Managers can apply these insights to enhance leadership practices and reduce burnout impact.

**Contribution/Value-Add:** This study clarifies the connection between servant leadership, burnout, and work ethic culture. The findings offer insights for future research and practical actions in organizational leadership.

**Keywords:** Servant Leadership, Burnout, Work Ethic Culture.

## Introduction

One crucial factor in developing the quality of human resources within an organization or company is leadership. According to (Smylie et al., 2002), the leadership used in an era of specialization and solely profit pursuit (often in the short term) is now deemed inappropriate and unworthy for use in the knowledge and integration era. A new approach in leadership is required, one that can simultaneously enhance the personal growth of employees and improve the quality and service of the institution. This involves the personal involvement of every member of the organization in decision-making processes and ethical and responsible behavior. Therefore, a new approach in the world of leadership is needed. Leader involvement in efforts to improve work quality and employee behavior is implemented in a leadership model known as servant leadership (Eva et al., 2019).

Servant leadership offers an alternative style for organizational change. Research by (Kiker et al., 2019) shows that it is a unified and cohesive concept. Greenleaf (J. Wu et al., 2021) key emphasis is on serving subordinates, prioritizing their interests. (Lewis, 2019) characterizes servant leadership as primarily service-oriented, fostering positive relationships through respect, teamwork, and attentive listening. Implementing servant leadership in organizations aims to boost employee performance, as found in research by (Canavesi & Minelli, 2022). They highlight servant leadership's distinctiveness, emphasizing the priority of serving over leading in this model.

challenge of being a servant leader lies in initiating changes in attitude, perspective, and behavior. Leadership involves providing opportunities for subordinates to succeed, with a commitment to assisting them. A good leader believes in this integral part of their attitude and behavior. Tangible impacts of changes elevate subordinates' spirit and morale, contributing to overall productivity. Subordinates realize they receive deserved treatment, leading to improved performance in line with changing leadership behavior at all levels of the company (Maamari & Saheb, 2018).

(C.-H. Wu & Parker, 2017) stresses leaders' role in providing meaning to employees' work. Leadership crises contribute to burnout, affecting employee well-being and incurring organizational costs to address stress. This underscores the need for people-oriented leadership (Giorgi et al., 2018) findings reveal burnout's strong impact on performance, indicating a decline without timely intervention. In contrast, (Gong et al., 2019) suggest that, despite burnout in certain roles, nurses' self-control and creative skills mitigate its impact on performance. The study highlights the importance of addressing burnout promptly for sustained employee well-being and performance improvement.

This study explores challenges in transforming organizations in Bandung City's hospitality industry, West Java, Indonesia. Firstly, the industry grapples with employee burnout due to work pressure, impacting well-being. Secondly, organizational change faces resistance, especially in altering work culture and leadership. Thirdly, implementing servant leadership encounters challenges amid diverse cultures, stressing the need for integration. Fourthly, altering work ethic culture is vital, requiring effective strategies. Despite challenges, the potential positive impact of transformation and servant leadership is evident in efforts to enhance employee well-being, aiming for increased productivity. This research aims to understand the intricate interaction between servant leadership, work ethic culture, and the mediating influence of burnout.

## Literature review
### Concept of work ethic culture
Work Ethic Culture serves as the moral foundation in an organization, guiding behavior through shared values and beliefs (Adams & Rau, 2004). It encompasses integrity, honesty, and responsibility, promoting ethical interactions. According to The Ethics and Compliance Initiative (ECI), it is a set of values and norms guiding interactions within and outside the organization. Leadership, as emphasized by (Metwally et al., 2019), plays a vital role in shaping a strong Work Ethic Culture. Leaders set positive examples and establish high ethical standards for all members. Neglecting ethics, disrespecting rules, or non-compliance, as warned by Ronald Reagan, can lead to organizational downfall. (Becker, 2019) notes that a robust Work Ethic Culture enhances employee satisfaction, organizational performance, and fosters innovation and sustainability. It goes beyond formal ethical norms, actively shaping an organization's character and reputation.

The measurement of Work Ethic Culture involves a comprehensive evaluation of elements shaping ethical character within an organization. This assessment encompasses key dimensions reflecting commitment and ethical behavior across all organizational activities. As outlined by (Becker, 2019), essential dimensions in measuring Work Ethic Culture include: (1) Organizational Integrity, gauging the extent to which the organization upholds values of integrity in all actions, demonstrating steadfastness to moral and ethical principles; (2) Honesty and Transparency, measuring the level of truthfulness in communication and the organization's openness regarding decisions and decision-making processes; (3) Social and Environmental Responsibility, evaluating the organization's commitment to positive societal and environmental impact; (4) Equality and Fairness, assessing the organization's promotion of equality and fairness in policies and actions, ensuring equitable treatment for all members; (5) Compliance and Ethical Behavior, evaluating the organization's adherence to rules, regulations, and ethical conduct in various contexts; (6) Employee Empowerment, measuring the organization's efforts to empower employees by providing freedom and support for ethical actions; (7) Employee Well-being, evaluating the organization's attention to the physical, mental, and emotional well-being of employees; and (8) Reputation and Image, highlighting how the organization is perceived in terms of integrity and ethics by the public, employees, and other stakeholders.

### Concept of servant leadership
Servant leadership places the leader as the primary servant, dedicated to the well-being and development of their team Greenleaf (J. Wu et al., 2021). It prioritizes serving subordinates and creating a values-based work climate. This approach centers decision-making on service, fostering relationships through dignity, respect, and community building. (Lewis, 2019). Servant leaders prioritize subordinates' needs and development (Eva et al., 2019), leading to satisfaction and improved organizational performance (Canavesi & Minelli, 2022). By embracing servant leadership principles, organizations create an environment supporting personal growth and productive collaboration. Recognized across sectors, servant leadership is not just an alternative but a paradigm shift in viewing a leader's role for collective success and individual growth.

Servant Leadership stands out as a distinctive leadership approach that not only involves managerial duties but also nurtures a robust and sustainable Work Ethic Culture. In this perspective, leaders transform from authoritative figures to dedicated servants, prioritizing the needs and growth of their team. Greenleaf (J. Wu et al., 2021) notes that Servant Leadership establishes an ethical environment consistently applying moral values. This concept positively influences Work Ethic Culture, emphasizing honesty, transparency, and care for team members. (Lewis, 2019) explains that Servant Leadership principles shape a culture of dignity, respect, and collaboration. By prioritizing subordinates' interests, servant leaders naturally foster an environment supportive of ethical values. (Sullivan, 2019) highlights that Work Ethic Culture shaped by servant leadership enhances employee satisfaction, boosts organizational performance, and encourages an innovative climate. Servant leaders, focusing on service, not only guide subordinates toward organizational goals but also inspire a culture that values ethics. Through Servant Leadership, organizations can cultivate a Work Ethic Culture that reinforces integrity, responsibility, and collaboration across all levels, making leaders catalysts for positive organizational change.

The measurement of Servant Leadership assesses how well servant leadership principles are applied and impactful in an organizational context. It includes dimensions reflecting key servant leader characteristics. According to Greenleaf (J. Wu et al., 2021), these dimensions encompass Service, recognizing a leader's dedication to others by actively meeting their needs. Another vital dimension is Individual Development, highlighting the leader's role in supporting subordinates' personal growth. Servant leadership also involves Community Building, evaluating a leader's ability to create a collaborative work environment. Empathetic Listening is crucial, assessing a leader's capacity to listen attentively and understand subordinates' perspectives. Lastly, servant leaders should demonstrate high Awareness of team and organizational dynamics. Measurement involves evaluating a leader's awareness of needs and challenges in their surroundings.

## Concept of Burnout

The World Health Organization defines burnout as chronic, untreated work-related stress leading to exhaustion, negative work attitude, and declining performance. Maslach and Leiter identify emotional exhaustion, depersonalization, and lack of personal accomplishment as key dimensions (Wanzhi, 2020). Addressing burnout requires analyzing risk factors and prevention strategies (Rożnowski, 2021). Proactive efforts can create a supportive work environment, crucial for both individual well-being and organizational sustainability. Understanding burnout is essential in navigating the challenges of the modern workplace, fostering a healthy and resilient work environment.

Burnout, a complex and widespread issue, significantly influences an organization's Work Ethic Culture. It alters the work environment dynamics, impacts individual behavior, and jeopardizes ethical integrity. Burnout is not just an individual concern; it affects organizational ethics and norms. Understanding this link is vital because burnout leads to negative attitudes, disengagement, and decreased motivation, impacting interactions and ethical norms. Maslach and Leiter's research (Wanzhi, 2020) emphasizes burnout's role in creating an unhealthy work environment and damaging interpersonal relationships. Depersonalization in burnout harms positive Work Ethic Culture. (Rożnowski, 2021) stresses organizations recognizing burnout's threat to cherished ethics. Maintaining a robust Work Ethic Culture involves addressing employees' mental well-being and preventing burnout. Managing burnout and preserving a positive Work Ethic Culture necessitate recognizing burnout signs, providing mental health support, and implementing preventive measures. A deep understanding of the burnout and Work Ethic Culture relationship is vital for fostering a healthy, ethical work environment.

Maslach and Leiter in (Wanzhi, 2020) present the Maslach Burnout Inventory (MBI), a tool measuring burnout through three main dimensions: (1) Emotional Exhaustion, indicating emotional fatigue, stress, and exhaustion; (2) Depersonalization, linked to a negative and cynical attitude toward work or individuals, manifesting as indifference or loss of humanity in interactions; and (3) Reduced Personal Accomplishment, measuring decreased success and personal effectiveness. The MBI serves as a widely used standard tool for assessing work-related burnout, combining these dimensions to identify and measure its presence in individuals.

## Concept of burnout mediates the relationship between servant leadership and work ethic culture

The concept of burnout plays a significant role in connecting Servant Leadership with the Work Ethic Culture in the workplace. In this relationship, burnout serves as a mediation that can illustrate how the influence of servant leadership can affect the dynamics of the work ethic culture within the organization. Servant Leadership, with its focus on service and the development of team members, can be a key factor in preventing or minimizing the level of burnout among employees. When leaders show attention and concern for the well-being of subordinates, it can help reduce factors that may lead to burnout, such as excessive workload or lack of social support.

A study by (Eva et al., 2019) shows that servant leadership reduces employee burnout, fostering a positive work environment. Servant leadership balances job demands and personal needs, preventing burnout. Burnout serves as a mediator between servant leadership and forming a strong Work Ethic Culture. Assessing burnout levels provides insights into how servant leadership impacts workplace ethics. (J. Wu et al., 2021) highlight the mediator role of burnout, aiding organizations in focusing on areas to enhance work ethic culture. Understanding the complex relationship between Servant Leadership, burnout, and Work Ethic Culture offers valuable insights for effective organizational management.

## Hypothesis

In this research, it is assumed that aspects of servant leadership positively impact work ethic culture. Leaders prioritizing service and caring for subordinates tend to foster a work environment rooted in ethical values, dedication, and professionalism. This aligns with (Greenleaf, 2019), emphasizing that servant leadership encourages service, building a work culture grounded in integrity and responsibility. However, the research hypothesis suggests that burnout, or high emotional fatigue, may mediate this relationship. As per Maslach and Leiter (Wanzhi, 2020), burnout not only causes dissatisfaction and exhaustion individually but can also impact overall organizational culture. High fatigue can affect employee enthusiasm and motivation, making it a relevant mediator between servant leadership and work ethic culture. This research aims to explore the positive influence of servant leadership on work ethic culture and the mediating role of burnout. According to (Demirtas & Akdogan, 2015), such research contributes significantly to understanding the intricate interaction between key factors in leadership and organizational culture. The conceptual framework of this research hypothesis is illustrated in the following figure:

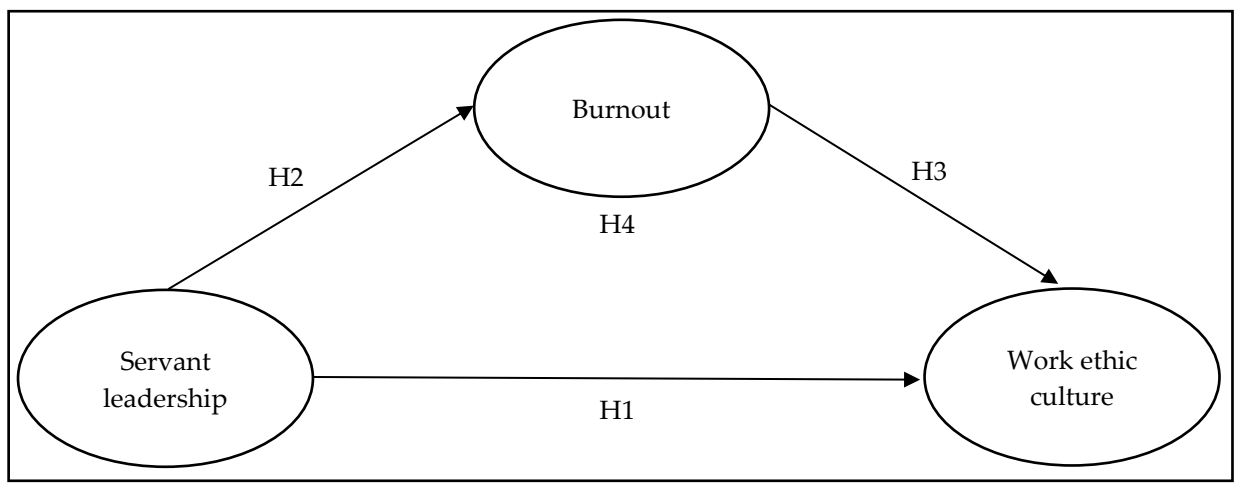

**Figure 1.** Conceptual Framework of the Study

H1 : There is an influence of servant leadership on ethical work culture.
H2 : There is an influence of servant leadership on burnout.
H3 : There is an influence of burnout on ethical work culture.
H4 : Burnout is capable of mediating the influence of servant leadership on burnout.

# Research method

## Design

The approach used in this research is quantitative with a survey method. The quantitative approach using a survey method is a research strategy that focuses on collecting structured data with the aim of measuring and analyzing relationships between variables. This method utilizes instruments such as questionnaires to gather data from representative respondents. The quantitative approach is often employed to achieve generalizations and provide a broader overview of the phenomena under investigation. By adopting this approach, the research can generate measurable data that can be statistically analyzed, providing a basis for the generalization of findings and supporting the validity and reliability of the research. In the words of (Creswell John and Creswell David, 2023), " The quantitative approach is used to identify relationships between variables, test theories, and develop and test models."

## Data collection

The data collection for this research involved conducting surveys in 4 and 5-star hotels in the city of Bandung, West Java Province, Indonesia. The study targeted a population of 113 hotels, involving a total of 339 employees, and data collection took place from September to December 2023. The sample consisted of 183 respondents, selected using the Slovin formula with a margin of error set at 0.05. The sampling technique used was Area Probability sampling, specifically employing purposive sampling to ensure that selected respondents reflected the desired characteristics for the study. This method enhances the relevance and applicability of findings to the specific context of the hospitality industry in the designated region. As expressed by (Neuman, 2014), "Survey research is a powerful method for gaining a broad view of a research issue, especially when studying large groups." In this regard, the survey approach played a significant role in collecting quantitative data from a considerable number of hotel employees, enabling a comprehensive analysis of the relationships between servant leadership, ethical work culture, and burnout in the hospitality sector. Furthermore, the use of the Slovin formula for determining the sample size and the application of the Area Probability sampling method demonstrate a systematic and robust statistical approach to ensure the representation of the target population in this research, aligning with best practices in survey research (Creswell John and Creswell David, 2023).

## Measurements

This research uses a comprehensive approach to measure its main variables. Servant leadership dimensions, including Service, Individual Development, Community Building, Empathic Listening, and Awareness, are assessed using indicators by Greenleaf (J. Wu et al., 2021). Burnout levels are measured with indicators proposed by Maslach and Leiter, detailed in research by (Wanzhi, 2020), covering Emotional Exhaustion, Depersonalization, and Reduced Personal Accomplishment. Ethical Work Culture is gauged using indicators by (Becker, 2019), encompassing Organizational Integrity, Honesty and Transparency, Social and Environmental Responsibility, Equality and Justice, Compliance and Ethical Behavior, Employee Empowerment, Employee Well-being, and Reputation and Image. This meticulous approach ensures accurate measurement of crucial aspects, forming a robust foundation for analyzing the impact of servant leadership and burnout on ethical work culture in the Bandung hospitality industry, West Java, Indonesia.

## Data analysis

In processing the collected data, this research applies the Structural Equation Model-Partial Least Square (SEM-PLS) methodology for analysis. This approach enables researchers to evaluate the complex relationships between variables involved in this study, particularly the influence of servant leadership on ethical work culture with burnout as a mediating factor. The SEM-PLS data analysis is chosen because of its advantages in handling complex models and non-normally distributed data. As expressed by (Schuberth et al., 2023), "SEM-PLS is suitable for research requiring predictive, exploratory, or confirmatory models, and it can address asymmetry and kurtosis in data." In a similar study, (Eva et al., 2019) affirmed that SEM-PLS is suitable for testing causal relationships among variables by accommodating models involving latent and manifest variables. By employing this approach, this study can detail the contribution of each variable and identify how servant leadership influences ethical work culture and the impact of burnout on that relationship. Thus, the use of SEM-PLS in the data analysis of this research reflects a commitment to a robust and relevant analytical approach to address complex research questions.

## Ethical considerations

In this study, ethics is a paramount consideration, adhering strictly to strong principles at every stage. Research ethics emphasize fair treatment, privacy rights, and participant safety. Ethical approval was obtained from the Institutional Review Board of Ariyanti Academy of Secretarial and Management, Bandung, Indonesia, prior to research commencement. Following Beauchamp and Childress (Bifarin & Stonehouse, 2022), research ethics includes principles like respecting autonomy, maintaining confidentiality, ensuring fairness, and balancing benefits and risks. This study is founded on a deep understanding of ethical implications, ensuring scholarly contribution with integrity and ethical prudence.

# Results and discussion
## Data analysis
### Measurement model

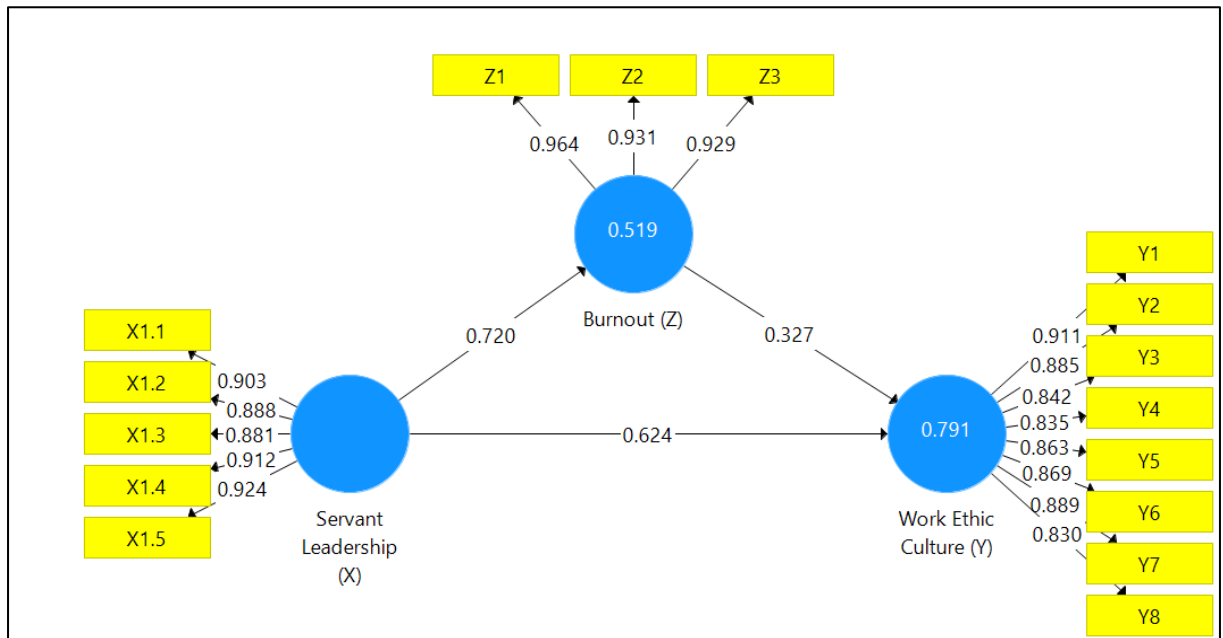

**FIGURE 1.** Outer Model

### Convergent validity

The measurement model explains how manifest variables represent the latent variables intended to be measured. Convergent validity measurement is evaluated through factor loading parameters, and AVE (Hair, 2021) states that an indicator has good correlation if the factor loading and AVE values exceed 0.7 with the measured construct.

**TABLE 1.** Summary of loading factor values

| Exogenous Construct | Dimension | Item Code | Loading Factor |
|---|---|---|---|
| Servant Leadership | Service | X.1 | 0.903 |
| | Individual Development | X.2 | 0.888 |
| | Community Building | X.3 | 0.881 |
| | Empathetic Listening | X.4 | 0.912 |
| | Awareness | X.5 | 0.924 |
| Burnout | Emotional Exhaustion | Z.1 | 0.964 |
| | Depersonalization | Z.2 | 0.931 |
| | Personal Accomplishment | Z.3 | 0.929 |
| Work Ethic Culture | Organizational Integrity | Y.1 | 0.911 |
| | Honesty and Transparency | Y.2 | 0.885 |
| | Social and Environmental Responsibility | Y.3 | 0.842 |
| | Equality and Justice | Y.4 | 0.835 |
| | Compliance and Ethical Behavior | Y.5 | 0.863 |
| | Employee Empowerment | Y.6 | 0.869 |
| | Employee Well-being | Y.7 | 0.889 |
| | Reputation and Image | Y.8 | 0.830 |

Table 1 above displays the factor loading values for all tested variables, indicating that all factor loading values are > 0.70, which suggests that we can proceed to the next testing phase.

## Discriminant validity

In conducting factor analysis or partial least squares structural equation modeling (PLS-SEM), discriminant validity testing is a crucial step. Examining cross-loadings, where indicators are expected to have higher factor loadings on their intended constructs compared to other constructs, is one commonly used technique. An additional metric to assess how well indicators capture the variance of the constructs is the Average Variance Extracted, or AVE. Higher AVE values indicate how well the indicators represent that particular component.

**TABLE 2**. Summary of cross loading

|     | Servant Leadership (X) | Burnout (Z) | Work Ethic Culture (Y) |
| --- | --- | --- | --- |
| X.1 | **0.903** | 0.681 | 0.815 |
| X.2 | **0.888** | 0.625 | 0.742 |
| X.3 | **0.881** | 0.712 | 0.738 |
| X.4 | **0.912** | 0.615 | 0.789 |
| X.5 | **0.924** | 0.608 | 0.791 |
| Z1 | 0.710 | **0.964** | 0.771 |
| Z2 | 0.653 | **0.931** | 0.681 |
| Z3 | 0.668 | **0.929** | 0.738 |
| Y1 | 0.792 | 0.765 | **0.911** |
| Y2 | 0.741 | 0.727 | **0.885** |
| Y3 | 0.738 | 0.671 | **0.842** |
| Y4 | 0.709 | 0.666 | **0.835** |
| Y5 | 0.744 | 0.755 | **0.863** |
| Y6 | 0.713 | 0.603 | **0.869** |
| Y7 | 0.768 | 0.614 | **0.889** |
| Y8 | 0.749 | 0.562 | **0.830** |

## Composite reliability

In testing the reliability of a construct, two commonly used methods are Cronbach's Alpha and Composite Reliability (CR). Cronbach's Alpha measures the internal consistency among items within a construct and is considered adequate if its value exceeds 0.7. Meanwhile, CR assesses the consistency among items by considering factor loadings, and CR values are considered good if they exceed 0.7 or 0.8. The use of both methods together can provide a more comprehensive understanding of construct reliability, with Cronbach's Alpha offering insights into the internal consistency of items, and CR providing a more sophisticated view by incorporating factor loadings. Together, they work to ensure that the constructs measured in a study are reliable and consistent, assisting researchers in interpreting the results of Structural Equation Modeling (SEM) or Partial Least Squares Structural Equation Modeling (PLS-SEM) analyses.

**TABLE 3**. Construct Reliability and Validity

|  | Cronbach's Alpha | Composite Reliability | Average Variance Extracted (AVE) |
| --- | --- | --- | --- |
| Servant Leadership (X) | 0.942 | 0.956 | 0.813 |
| Burnout (Z) | 0.936 | 0.959 | 0.886 |
| Work Ethic Culture (Y) | 0.952 | 0.960 | 0.750 |

## Higher order confirmatory factor analysis (HCFA)

The second-order confirmatory factor analysis, or Higher Order Confirmatory Factor Analysis (HCFA), is a valuable method for assessing the theoretical relationships between higher-order categories and more specific constituents (Hair, 2021).

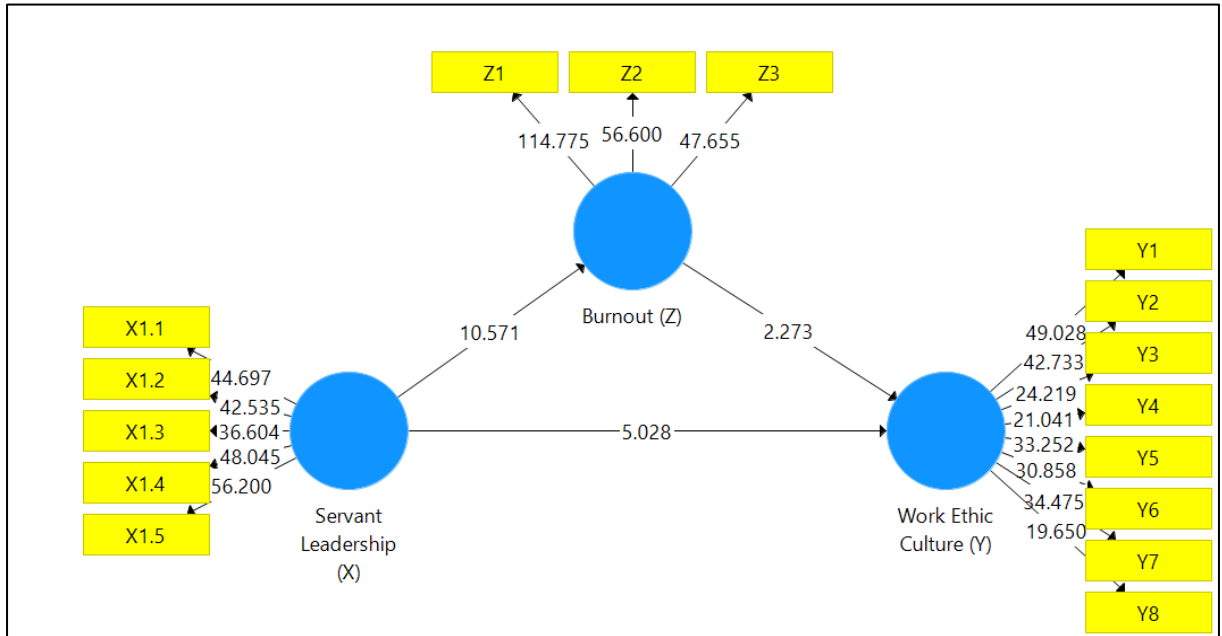

**FIGURE 2.** Inner Model

The bootstrap process is employed to estimate confidence intervals, test statistical significance, or validate a model. The t-values from bootstrap for a two-tailed test of 1.96 reflect the commonly used threshold in statistics to determine the significance of a value. In this context, a t-value of 1.96 at a significance level of 0.05 corresponds to the 0.05 significance level in the t-distribution.

**TABLE 4.** Significance of measurement path coefficients in SCFA

| | Original Sample (O) | Sample Mean (M) | Standard Deviation (STDEV) | T Statistics (|O/STDEV|) | P Values |
|---|---|---|---|---|---|
| Servant Leadership (X) -> Work Ethic Culture (Y) | 0.624 | 0.599 | 0.124 | 5.028 | 0.000 |
| Servant Leadership (X) -> Burnout (Z) | 0.720 | 0.719 | 0.068 | 10.571 | 0.000 |
| Burnout (Z) -> Work Ethic Culture (Y) | 0.327 | 0.350 | 0.144 | 2.273 | 0.023 |

## Structural model analysis

The Structural Model Analysis in PLS-SEM is a statistical method used to test and measure cause-and-effect relationships in a complex theoretical model. Structural Model Analysis in the context of Partial Least Squares Structural Equation Modeling (PLS-SEM) involves the use of R-squared ($R^2$) to assess the extent to which the model explains variation in endogenous variables. R-squared is utilized to measure the success of the model in explaining the variability of the target variable or the final variable in a structural model.

## R-square ($R^2$)

**TABLE 5.** R-square ($R^2$)

| | R Square | R Square Adjusted |
|---|---|---|
| Burnout (Z) | 0.519 | 0.516 |
| Work Ethic Culture (Y) | 0.791 | 0.789 |

The model regarding the effect of servant leadership on burnout has a value of 0.519, as seen from the provided data. It can be understood that the burnout construct contributes 51.9% to the variability of the organizational ethical culture construct, with other factors not covered in this study contributing to the remaining variability. Similarly, the value of 0.935 is the value of the model analyzing how servant leadership influences organizational ethical culture. It can be understood that the variability in the servant leadership construct contributes approximately 79.1% to the variability of the organizational ethical culture construct, with the remaining variability explained by other factors.

## Hypothesis testing
### The influence of servant leadership on work ethic culture

In Table 4, it is evident that the construct of servant leadership has a significant impact (β=0.624) on the construct of organizational ethical culture, with a t-statistic between their relationship at 5.028 > 1.96, and a p-value of 0.000 < 0.05. Thus, the first hypothesis stating that servant leadership influences organizational ethical culture is supported. These findings provide an insight that this leadership model signifies a deep commitment to serving others, creating a robust foundation for a strong ethical culture. Servant leaders, by setting an example through ethical behavior and decisions, stimulate the development of work norms that emphasize integrity, honesty, and responsibility. The success of servant leadership in creating an environment that prioritizes the well-being of team members, attends to individual needs, and encourages collaboration, all contribute to the formation of a positive ethical culture. Servant leadership, with its focus on serving others, has been recognized as a leadership model that can have a positive and significant impact on organizational ethical culture in various settings. In recent years, research and expert perspectives have provided a deeper understanding of the relationship between servant leadership and a strong ethical work culture. These results align with the research conducted by (Greenleaf, 2019) yang menemukan bahwa which found that servant leadership influences organizational ethical culture. The findings of this study also support previous research stating that servant leadership influences organizational ethical culture (Davis, 2018), (Aqli et al., 2019), (Brown et al., 2020), and (Bush, 2023).

### The influence of servant leadership on burnout

In Table 4, it can be observed that the construct of servant leadership has a significant impact ($\beta$=0.327) on the construct of burnout, with a t-statistic between their relationship at 2.273 > 1.96, and a p-value of 0.023 < 0.05. Thus, the second hypothesis stating that servant leadership influences burnout is supported. These findings provide an insight that the servant leadership model has a positive and significant impact in reducing or preventing burnout in the workplace. Servant leadership, which focuses on serving others, individual development, and creating a supportive work environment, is considered an effective approach to managing stress and workload that can lead to burnout. By applying the principles of servant leadership, leaders strive to understand and meet the needs of team members, ensure a healthy work-life balance, and create an organizational culture that supports employee well-being. Servant leadership is regarded as a solution that can make a substantial positive contribution to managing burnout in the workplace. Recent research and expert insights provide a deeper understanding of how servant leadership can be an effective solution for addressing burnout. These results align with the research conducted by (Feng & Adams, 2023), which found that servant leadership influences burnout. The findings of this study also support previous research stating that servant leadership influences burnout (Acta, 2021), (Miller et al., 2022), and (Di Fabio & Peiró, 2023).

### The influence of burnout on work ethic culture

In Table 4, it is evident that the construct of burnout has a significant impact ($\beta$=0.720) on the construct of ethical work culture, with a t-statistic between their relationship at 10.571 > 1.96, and a p-value of 0.000 < 0.05. Thus, the third hypothesis stating that burnout influences ethical work culture is supported. These findings provide an insight that the relationship between burnout and ethical work culture is complex and can have a significant impact on organizational dynamics. Burnout, as a form of physical and mental fatigue caused by chronic work stress, can undermine the foundation of a healthy ethical culture. Employees experiencing burnout tend to show decreased motivation, emotional fatigue, and a lack of engagement in their work. This, in turn, can lead to unethical behavior and neglect of the ethical values that should shape the work culture. Steps to manage stress, support employees' mental well-being, and promote a healthy work-life balance become essential to mitigate the negative impact of burnout and ensure the sustainability of a positive ethical culture. In a recent study, (Johnson et al., 2022) stated, "Burnout is significantly related to a decline in the ethical work culture within organizations. Employees experiencing emotional exhaustion are likely to be less concerned about ethical values and may be more susceptible to unethical behavior in the workplace." This reflects how burnout can undermine the foundation of a healthy ethical culture, which should be based on integrity, honesty, and responsibility.

### Testing mediation effects

Using Partial Least Squares (PLS), examining the relationships among independent, dependent, and mediating variables is a key component in Structural Equation Modeling analysis (PLS-SEM) with mediating effects. When a mediating variable partially or fully explains the relationship between an independent variable and a dependent variable, mediation occurs. Three prerequisites must be met to test the mediation effect (Creswell John and Creswell David, 2023). The first requirement is that there must be a significant relationship between the independent variable and the dependent variable overall. The t-statistic value, which should be >1.96 to indicate significance at a 95% confidence level, can be used to test this research. The substantial impact of the independent variable on the mediating variable is the second requirement. The t-statistic value, which should also be >1.96, is used to test this research. The strong impact of the mediating variable on the dependent variable is the third requirement. Further testing for this involves looking at the t-statistic value, which should be more than 1.96. At this point, it is anticipated that the mediating variable will continue to have a significant impact on the dependent variable while the main effect becomes insignificant. It can be concluded that there is a significant mediation effect if all three prerequisites are met. According to (Creswell John and Creswell David, 2023), this indicates that the relationship between the independent and dependent variables is mediated by the mediating variable. Servant leadership significantly influences ethical work culture in the first stage (t-statistic 5.028 > 1.96), as shown in Table 4. The t-statistic 2.135 > 1.96. Servant leadership significantly influences burnout in the second stage (t-statistic 10.571 > 1.96), as indicated in Table 4, and the study moves to the next phase. The third stage involves the simultaneous examination of the mediating variable, burnout, and the independent variable, servant leadership, on the dependent variable, ethical work culture.

**TABLE 6.** Total effect

|  | Original Sample (O) | Sample Mean (M) | Standard Deviation (STDEV) | T Statistics (\|O/STDEV\|) | P Values |
|---|---|---|---|---|---|
| Burnout (Z) -> Work Ethic Culture (Y) | 0.327 | 0.350 | 0.144 | 2.273 | 0.023 |
| Servant Leadership (X) -> Burnout (Z) | 0.720 | 0.719 | 0.068 | 10.571 | 0.000 |
| Servant Leadership (X) -> Work Ethic Culture (Y) | 0.860 | 0.857 | 0.034 | 25.337 | 0.000 |

Based on the results of the PLS analysis above, it is evident that ethical work culture is significantly influenced by burnout (β=0.327), with a t-statistic of 2.273 surpassing the threshold of 1.96. Servant leadership has a significant impact on burnout (β=0.720), as indicated by a t-statistic of 10.571, exceeding the 1.96 criteria. With a t-statistic of 25.337 and β=0.860, servant leadership has a significant impact on ethical work culture above the 1.96 criteria. Therefore, it can be concluded that the fourth hypothesis stating that servant leadership and ethical work culture are mediated by burnout is confirmed.

**TABLE 7.** Specific indirect effect

|  | Original Sample (O) | Sample Mean (M) | Standard Deviation (STDEV) | T Statistics (\|O/STDEV\|) | P Values |
|---|---|---|---|---|---|
| Servant Leadership (X) -> Burnout (Z) -> Work Ethic Culture (Y) | 0.236 | 0.258 | 0.131 | 2.189 | 0.021 |

We can refer to the table of specific indirect effects to determine the extent to which burnout can moderate the relationship between servant leadership and ethical work culture. The findings of the analysis indicate that, with a t-statistic of 2.189, surpassing the 1.96 criteria, the relationship between servant leadership and ethical work culture, mediated by burnout, remains significant. According to (Hair, 2021), this indicates that burnout serves as a partial control, demonstrating the presence of both direct and indirect effects among the variables.

The link between servant leadership and ethical work culture is vital for creating a healthy and integrity-driven organizational environment. A servant leader, prioritizing service and attention to team members' needs, can be a catalyst for a positive ethical culture. However, this relationship is not always straightforward and can be affected by external factors, including team members' burnout. Burnout becomes a significant obstacle to the effectiveness of servant leadership in building a robust ethical work culture. Team members experiencing burnout often exhibit decreased motivation and engagement, undermining the foundation of the ethical culture. As noted by (Green & Kinchen, 2021), 'Burnout can challenge servant leaders in maintaining engagement and commitment to ethical values amid prolonged work stress.'

Nevertheless, servant leadership remains crucial in alleviating burnout and fortifying the ethical work culture. As highlighted by (Johnson et al., 2022), "Servant leaders, prioritizing service and the well-being of team members, can create an environment where burnout is identified and addressed, supporting a positive ethical culture." To tackle burnout's impact, servant leaders can implement practices like guiding individuals to achieve a healthy work-life balance and ensuring that ethical values underpin every decision. Despite the challenge posed by burnout, servant leadership retains the potential to positively influence ethical work culture through its dedication to service, care, and individual development.

# Conclusion and recommendation

In summary, this study establishes a noteworthy connection among servant leadership, burnout, and ethical work culture in the hospitality industry. Servant leadership significantly and positively influences ethical work culture by fostering an environment of integrity and responsibility. Additionally, the study reveals that servant leadership also has a positive impact on burnout, underscoring the need for adept leadership to manage team members' physical and mental fatigue effectively, promoting a healthy work environment. Moreover, burnout is shown to significantly influence ethical work culture, indicating that fatigue levels can shape employees' behavior and attitudes towards ethical values. Lastly, the study demonstrates that burnout acts as a mediating factor in the relationship between servant leadership and ethical work culture, explaining how servant leadership influences ethical work culture through team members' fatigue levels. Recommendations stemming from these findings emphasize the recognition of servant leadership's pivotal role in fostering a positive ethical work culture in the hospitality industry. Implementing preventive measures against burnout, such as stress management programs and employee well-being support, is crucial. Organizations are encouraged to design leadership strategies that address individual needs, promote personal development, and actively cultivate an environment supportive of ethical values.

# Acknowledgements

In completing this research, we extend our gratitude to key contributors. Firstly, we acknowledge the support and guidance provided by the Academy of Secretarial and Management Ariyanti in Bandung, Indonesia. Their assistance proved invaluable throughout the research process. Secondly, we appreciate the willingness of General Managers, Managers, and Supervisors from 4 and 5-star hotels in Bandung, Indonesia, who participated as subjects in this research. Their collaboration significantly facilitated the study's progression. We also thank other contributors, both in moral support and technical assistance, who, though not individually named, played crucial roles in various aspects. The combined efforts of these contributors have greatly enriched the depth and significance of this research.

# Competing interests

In conducting this research, the authors affirm that there are no competing interests that could affect the integrity or objectivity of the research results.

# Authors' contributions

In this research, the authors made diverse contributions. The first author led the planning, design, data collection, analysis, and manuscript drafting. The second author concentrated on the literature review and theoretical framework. The third author contributed to data analysis and conclusions. The fourth and fifth authors participated in the editing and final formatting. Collaboratively, the authors played essential roles, combining their unique strengths to ensure the comprehensive and high-quality completion of this manuscript.

# Funding information

In this research, the authors extend gratitude to the funding source, the Academy of Secretary and Management Ariyanti in Bandung, Indonesia, for making this study possible. The financial support played a crucial role in facilitating research activities, data acquisition, analysis, and the drafting and publication of findings, contributing to the success and smooth implementation of the research.

# Data availability

In this study, the authors would like to emphasize that the data underlying the findings and results presented in this article are available to the public. All datasets, data collection methods, and research instruments used can be accessed by interested parties or other researchers who wish to examine or replicate this study

# Disclaimer

In this study, we emphasize that the presented results and perspectives are objective and derived from research findings. However, readers are reminded that any interpretation or use of this information is their personal responsibility. The authors are not accountable for errors or inaccuracies originating from external sources and do not guarantee the certainty or completeness of the research results.